\documentclass[runningheads]{svmult}

\usepackage{makeidx}   
\usepackage{graphicx}  
\usepackage{subeqnar}  
\usepackage{multicol}  
\usepackage{physprbb}  
\makeindex             

\usepackage{epsfig}
\usepackage{psfig}
\usepackage{eepic}
\usepackage{epic}
\usepackage{amsmath}
\usepackage{amssymb}  
\usepackage{latexsym}

\newcommand{\vm}{v_{\rm max}}

\begin{document}
\title*{Anomalous fundamental diagrams in traffic on ant trails}
\toctitle{Anomalous fundamental diagrams in traffic on ant trails}
%
%
\titlerunning{Anomalous fundamental diagrams in traffic on ant trails}
%
\author{Andreas Schadschneider\inst{1}
\and Debashish Chowdhury\inst{2}
\and Alexander\ John\inst{1}
\and Katsuhiro Nishinari\inst{3}}
\authorrunning{A.\ Schadschneider, D.~Chowdhury, A.\ John, K.\ Nishinari}
%
%
\institute{Institut f\"ur Theoretische Physik, Universit\"at zu K\"oln,
50937 K\"oln, Germany, {\tt [as,aj]@thp.uni-koeln.de}
\and
Department of Physics, 
Indian Institute of Technology,
Kanpur 208016, India, {\tt debch@iitk.ac.in} 
\and
Department of Applied Mathematics and Informatics,
Ryukoku University, Shiga 520-2194, Japan, {\tt knishi@rins.ryukoku.ac.jp}
}

\maketitle              

\begin{abstract}
Many insects like ants communicate chemically via chemotaxis. This allows 
them to build large trail systems which in many respects are similar to 
human-build highway networks. Using a recently proposed stochastic cellular 
automaton model we discuss the basic properties of the traffic flow on 
existing trails.
Surprisingly it is found that in certain regimes the average speed of 
the ants can vary non-monotonically with their density. This is in sharp
contrast to highway traffic. The observations can be
understood by the formation of loose clusters, i.e.\ space
regions of enhanced, but not maximal, density. We also discuss the
effect of counterflow on the trails.
\end{abstract}


\section{Introduction}
\label{sec_intro}

The occurance of organized traffic \cite{css,helbing,nagatani}
is not only restricted to human societies. Prominent examples 
of self-organized motion in biology are herding, flocking and
swarm formation \cite{herding,flocking,swarm,chemo}. In addition,
especially ants build trail systems that have many similarities
with highway networks \cite{franks01}.

Ants communicate with each other by dropping a chemical (generically
called {\it pheromone}) on the substrate as they move forward
\cite{wilson,camazine,mikhailov}. Although we cannot smell it, the
trail pheromone sticks to the substrate long enough for the other
following sniffing ants to pick up its smell and follow the trail.
This process is called {\em chemotaxis} \cite{chemo}.
Ant trails may serve different purposes (trunk trails, migratory
routes) and may also be used in a different way by different species.
Therefore one-way trails are observed as well as trails with
counterflow of ants.

In the following we will not discuss the process of trail formation
itself. This has been studied in the past (see e.g.\ \cite{schweitzer,couzin}
and references therein). 
Instead we assume the existence of a trail network that is constantly
reinforced by the ants. Focussing on one particular trail it is natural
to assume that the motion of the ants is one-dimensional.

In traffic flow interactions between the vehicles typically lead to 
a reduction of the average velocity $\bar{v}$. This is obvious for braking 
maneouvers to avoid crashes. Therefore $\bar{v}(\rho)$ decreases
with increasing density $\rho$.\footnote{A possible exception is a
synchronized phase where the correlation between current and density
is very small \cite{synchro}.} In the following we will show that this can 
be different for ant-trails. Even though due to the similar velocity of 
the ants overtaking is very rare, which implies that a description in terms
of an exclusion process is possible, the presence of the pheromone
effectively leads to an enhancement of the velocity.

We just briefly mention that similar ideas can be applied to the description
of pedestrian dynamics. In \cite{pede1,pede2} we have developed a 
pedestrian model based on virtual chemotaxis. The basic principles are
very similar to the ant-trail model introduced below. However, the
motion in pedestrian dynamics is essentially two-dimensional.


\section{Definition of the model}
\label{sec_model}

In \cite{cgns,ncs,jscn} we have developed a particle-hopping model, 
formulated in terms of a stochastic cellular automaton (CA), which may 
be interpreted as a model of uni-directional flow in an ant-trail. 
As mentioned in Sec.~\ref{sec_intro} we do not want to  address 
the question of the emergence of the 
ant-trail \cite{activewalker}, but focus on the traffic of 
ants on a trail which has already been formed. The model generalizes
the asymmetric simple exclusion process (ASEP) \cite{derrida,derrida2,schutz} 
with parallel dynamics by taking into account the effect of the pheromone.

The ASEP is one of the simplest examples of a system driven far from
equilibrium. Space is discretized into cells that can be occupied 
by at most one particle. In the totally asymmetric case (TASEP) particles 
are allowed to move in one direction only, e.g.\ to the right. If the 
right neighbour site is empty a particle hops there with probability $q$. 
For parallel (synchronous) dynamics this is identical to the limit 
$v_{\text{max}}=1$  of the Nagel-Schreckenberg (NaSch) model \cite{ns} 
with braking probability $p=1-q$. 

\begin{figure}[tb]
\begin{center}
\includegraphics[width=0.65\textwidth]{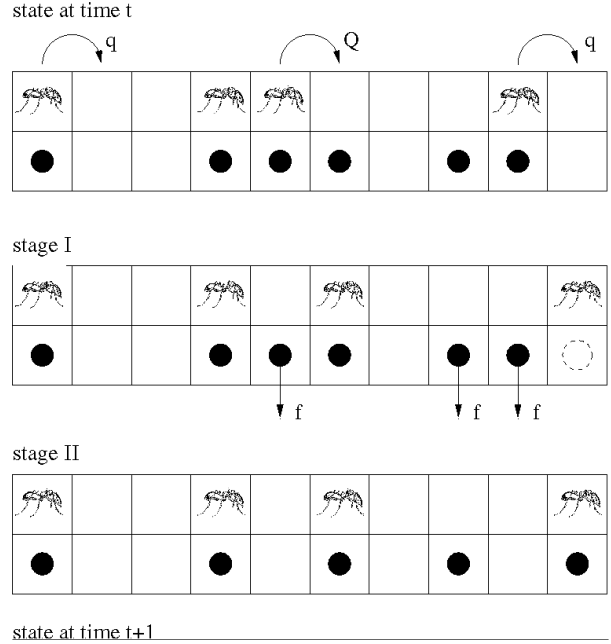} 
\end{center}
\caption{
Schematic representation of typical configurations of the 
uni-directional ant-traffic model. The symbols $\bullet$ indicate 
the presence of pheromone.
This figure also illustrates the update procedure. 
Top: Configuration at time $t$, i.e.\ {\it before} {\em stage I}
of the update. The non-vanishing probabilities of forward movement of 
the ants are also shown explicitly. Middle: Configuration {\it after} 
one possible realisation of {\it stage I}. Two ants have moved compared 
to the top part of the figure. The open circle with dashed boundary 
indicates the location where pheromone will be dropped by the corresponding 
ant at {\em stage II} of the update scheme. Also indicated are the existing 
pheromones that may evaporate in {\em stage II} of the updating, together 
with the average rate of evaporation.  Bottom: Configuration {\it after} 
one possible realization of {\it stage II}. Two drops of pheromones 
have evaporated and pheromones have been dropped/reinforced at the 
current locations of the ants.
}
\label{fig-unidir_model}
\end{figure}

In our model of uni-directional ant-traffic the ants move according to
a rule which is essentially an extension of the TASEP dynamics. 
In addition a second field is introduced which models 
the presence or absence of pheromones (see Fig.~\ref{fig-unidir_model}). 
The hopping probability of the ants is now modified by the presence of 
pheromones. It is larger if a pheromone is present at the destination site.
Furthermore the dynamics of the pheromones has to be specified. They
are created by ants and free pheromones evaporate with probability $f$.
Assuming periodic boundary conditions, the state of the system is updated at 
each time step in two stages (see Fig.~\ref{fig-unidir_model}). 
In stage I ants are allowed to move while in stage II the pheromones are 
allowed to evaporate. In each stage the {\it stochastic} dynamical rules 
are applied in parallel to all ants and pheromones, respectively.\\

\noindent {\it Stage I: Motion of ants}\\[0.2cm]
\noindent An ant in a cell cannot move if the cell immediately in front 
of it is also occupied by another ant. However, when this cell is not 
occupied by any other ant, 
the probability of its forward movement to the ant-free cell is $Q$ 
or $q$, depending on whether or not the target cell 
contains pheromone. Thus, $q$ (or $Q$) would be the average speed of a 
{\it free} ant in the absence (or presence) of pheromone. To be consistent 
with real ant-trails, we assume $ q < Q$, as presence of pheromone 
increases the average speed.\\

\noindent {\it Stage II: Evaporation of pheromones}\\[0.2cm]
\noindent Trail pheromone is volatile. So, pheromone secreted by an ant
will gradually decay unless reinforced by the following ants. In order to 
capture this process, we assume that each cell occupied by an ant at the 
end of stage I also contains pheromone. On the other hand, pheromone in 
any `ant-free' cell is allowed to evaporate; this evaporation is also 
assumed to be a random process that takes place at an average rate of $f$ 
per unit time. 

The total amount of pheromone on the trail can fluctuate although the 
total number $N$ of the ants is constant because of the 
periodic boundary conditions. In the two special cases $f = 0$ and 
$f = 1$ the stationary state of the model becomes identical to that of
the TASEP with hopping probability $Q$ and $q$, respectively. 

For a theoretical description of the process we associate two binary 
variables $S_i$ and $\sigma_i$ with each site $i$. $S_i$ is the
occupation number of ants and takes the value $0$ or $1$ depending on 
whether the cell $i$ is empty or occupied by an ant.
Similarly, $\sigma_i$ is the occupation number of the pheromones, i.e.\
$\sigma_i =  1$ if the cell $i$ contains pheromone; otherwise,
$\sigma_i =  0$. Thus, we have two subsets of dynamical variables in
this model, namely,
$\{S(t)\} \equiv \{S_1(t),S_2(t),...,S_i(t),...,S_L(t)\}$ 
and
$\{\sigma(t)\} \equiv \{\sigma_1(t),\sigma_2(t),...,\sigma_i(t),...,
\sigma_L(t)\}$.
In stage I the subset $\{S(t+1)\}$ at the time step $t+1$ is obtained 
using the full information $(\{S(t)\},\{\sigma(t)\})$ at time $t$. 
In stage II only the subset $\{\sigma(t)\}$ is updated so that at the 
end of stage II the new configuration $(\{S(t+1)\},\{\sigma(t+1)\})$ at 
time $t+1$ is obtained.

The rules can be written in a compact form as 
the coupled equations
\begin{eqnarray}
 S_j(t+1)&=&S_j(t)+\min\{\eta_{j-1}(t),S_{j-1}(t),1-S_j(t)\}\nonumber\\
&&\hspace{0.5cm}-\min\{\eta_{j}(t),S_{j}(t),1-S_{j+1}(t)\},\label{eqa}\\
\sigma_j(t+1)&=&\max\{S_j(t+1),\min\{\sigma_j(t),\xi_j(t)\}\},\label{eqf}
\end{eqnarray}
where $\xi$ and $\eta$ are stochastic variables defined by 
$\xi_j(t)=0$ with the probability $f$ and $\xi_j(t)=1$ with $1-f$, 
and $\eta_j(t)=1$ with the probability $p=q+(Q-q)\sigma_{j+1}(t)$ and 
$\eta_j(t)=0$ with $1-p$. This representation is useful for the
development of approximation schemes.


\section{Fundamental diagram for uni-directional motion}
\label{sec_fund}

\begin{figure}[tb]
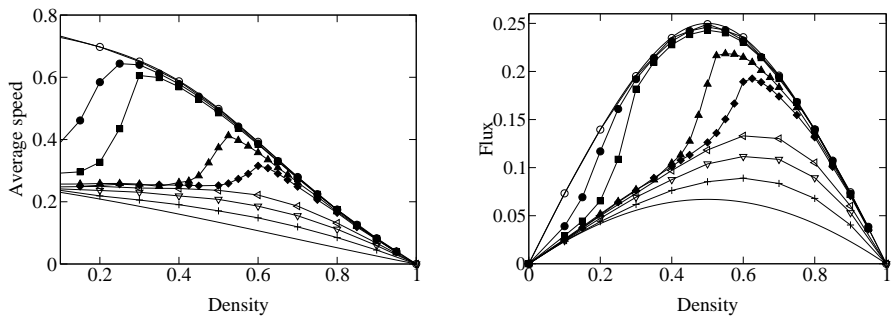

\begin{center}
\ \phantom{a}\\[0.2cm]
\includegraphics[width=0.45\textwidth]{schadfig2a.eps}
\qquad
\includegraphics[width=0.45\textwidth]{schadfig2b.eps}
\end{center}
\caption{ 
The average speed (left) and the flux (right) of the ants, in the 
{\it uni-directional} ant-traffic model, are plotted against
their densities for the parameters $Q = 0.75, q = 0.25$.
The curves correspond to $f=0.0001 ({\circ})$,
$0.0005 (\bullet$), $0.001 (\blacksquare)$, $0.005 (\blacktriangle)$ 
$0.01 ({\blacklozenge})$, $0.05 ({\triangleleft})$, 
$0.10 (\bigtriangledown)$, $0.25 ($+$)$. 
In both graphs, the cases $f=0$ and $f=1$ are also displayed by the 
uppermost and lowermost curves (without points); these are exact 
results corresponding to the TASEP or NaSch model with $\vm=1$
and hopping probability $Q$ and $q$, respectively. 
Curves plotted with filled symbols have unusual shapes.
}
\label{fig-unidir_res}
\end{figure}

In vehicular traffic, usually, the inter-vehicle interactions tend to 
hinder each other's motion so that the average speed of the 
vehicles decreases {\it monotonically} with increasing density.
This can be seen in Fig.~\ref{fig-unidir_res} where also
the fundamental diagram of the NaSch model with $v_{\text{max}}=1$ 
(or the TASEP with parallel updating) is shown for different hopping
probabilities. Note also the particle-hole symmetry of the flow
that satisfies $F_{\text{NS}}(\rho)=F_{\text{NS}}(1-\rho)$. Explicitly 
it is given by \cite{ssni}
\begin{equation}
F_{\text{NS}}(\rho)= \frac{1}{2}\left[1-\sqrt{1 ~- ~4 ~q_{\text{NS}} 
~\rho(1-\rho)}\right]
\label{eq-nsflux}
\end{equation}
where $q_{\text{NS}}=1-p$ with $p$ being the braking probability.
Here $\rho=N/L$ is the density of vehicles and $L$ the number of cells.

In contrast, in our model of uni-directional ant-traffic 
the average speed of the ants varies {\it non-monotonically} with their 
density over a wide range of small values of $f$ 
(see Fig.~\ref{fig-unidir_res}) because of the coupling of their 
dynamics with that of the pheromone. This uncommon variation of the 
average speed gives rise to the unusual dependence of the flux on the 
density of the ants in our uni-directional ant-traffic model 
(Fig.~\ref{fig-unidir_res}). Furthermore the flux is no longer 
particle-hole symmetric.

\subsection{\label{sec_loose}''Loose'' cluster approximation (LCA)} 

\begin{figure}[tb]
\begin{center}
\includegraphics[width=0.45\textwidth]{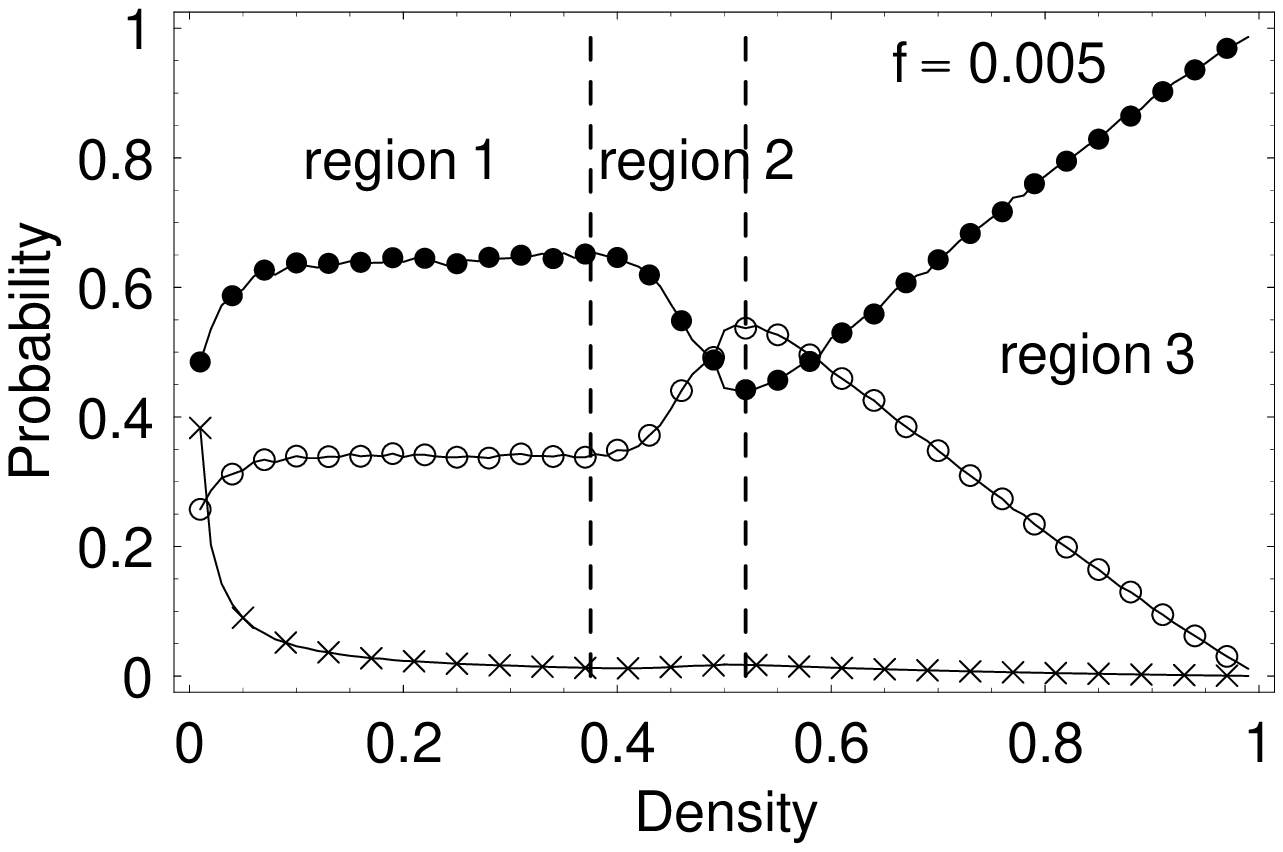}
\qquad
\includegraphics[width=0.45\textwidth]{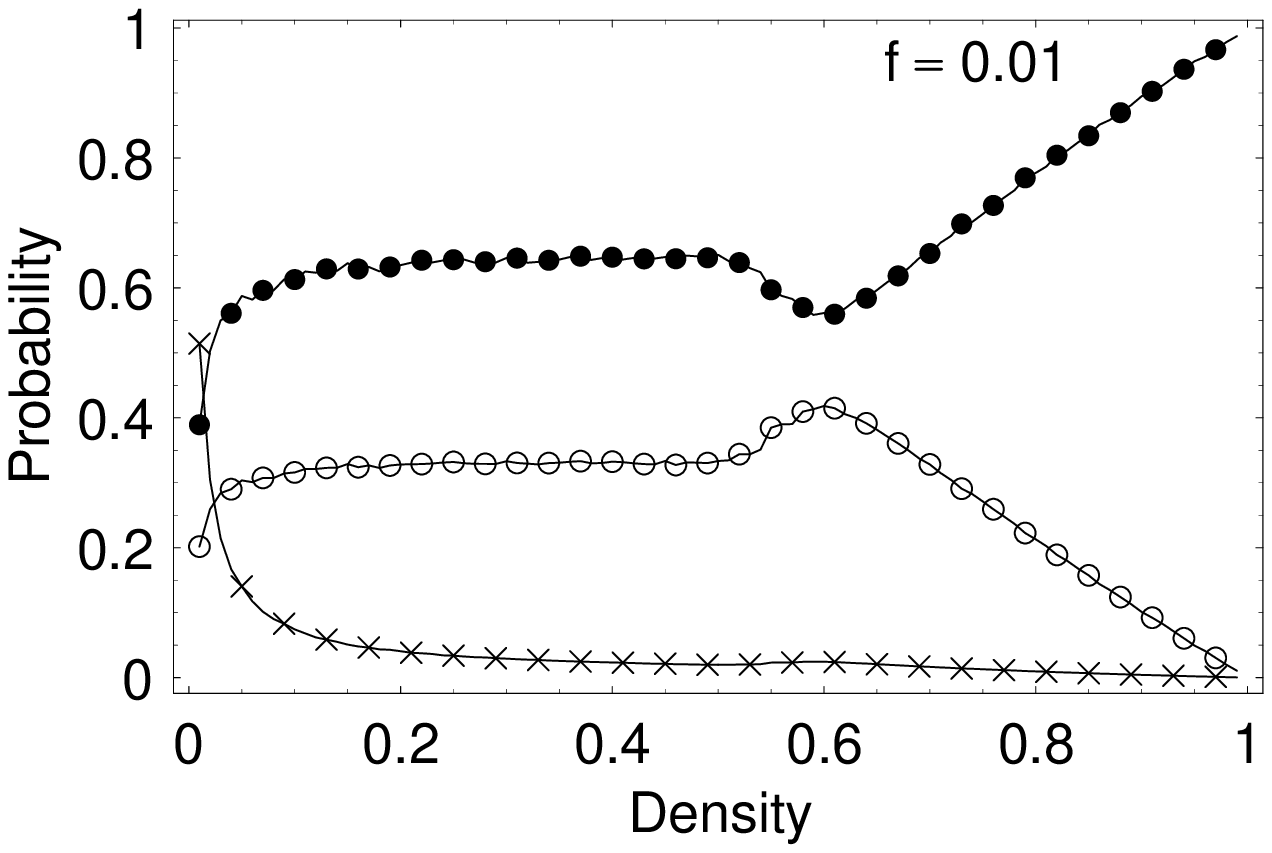}
\end{center}
\caption{Numerical results for the probabilities of finding an 
ant ($\bullet$), pheromone but no ant ($\circ$) 
and nothing ($\times$) in front of an ant are plotted against 
density of the ants.
The parameters are $f=0.005$ (left) and $f=0.01$ (right). 
}
\label{fig-lca}
\end{figure}

How can the unusual density-dependence of the average velocity
be understood? Using mean-field-type theories that
implicitly assume a homogeneous stationary state does not allow
to reproduce the simulation results in a satisfactory way \cite{ncs}.
This indicates that the stationary state is characterized by some
sort of clustering. This is confirmed by considering the probabilities 
of finding  an ant ($P_a$), pheromone ($P_p$) and nothing ($P_0$) in front 
of a cell occupied by an ant. Typical results from
computer simulations are shown in Fig.~\ref{fig-lca}. 

\begin{figure}[tb]
\begin{center}
\includegraphics[width=0.48\textwidth]{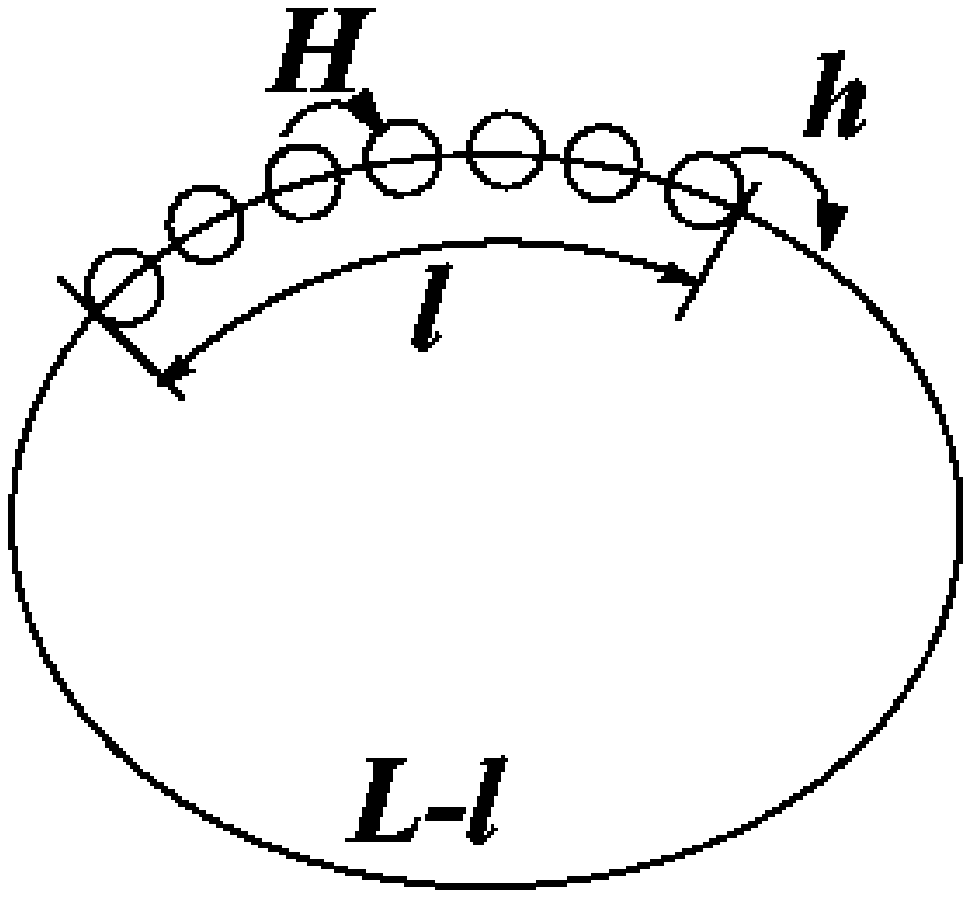} \quad
\includegraphics[width=0.45\textwidth]{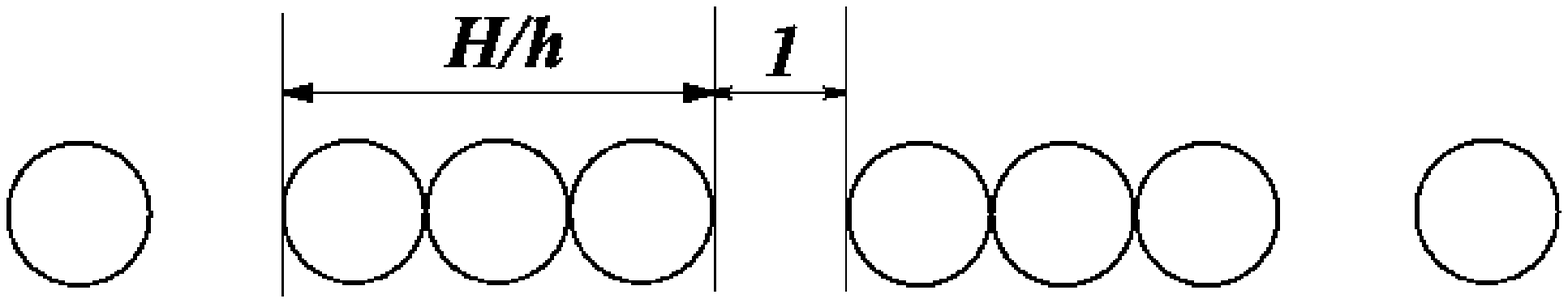}
\end{center}
\caption{(left) Schematic explanation of the loose cluster.
$H$ is the hopping probability of ants inside the loose cluster and
$h$ is that of the leading ant. (right) The stationary loose cluster.
The average gap between ants becomes $h/H$, 
which is irrelevant to the density of ants.
}
\label{fig-loose}
\end{figure}

One can distinguish three different regions. ``Region 1'' is
characterized by the flat part of the curves in Fig.~\ref{fig-lca} in
the low density regime. Here, in spite of low density of the ants, the
probability of finding an ant in front of another is quite high. This
implies the fact that ants tend to form a cluster.  On the other hand,
cluster-size distributions obtained from computer simulations
show that the probability of finding isolated ants are always higher than 
that of finding a cluster of ants occupying nearest-neighbor sites \cite{ncs}.
These two apparently contradictory observations can be reconciled by 
assuming that the ants form ``loose'' clusters in the region 1.  The 
term ``loose'' means that there are small gaps in between successive
ants in the cluster, and the cluster looks like an usual compact
cluster if it is seen from a distance (Fig.~\ref{fig-loose}). In other 
words, a loose cluster is just a loose assembly of isolated ants that
corresponds to a space region with density larger than the
average density $\rho$, but smaller than the maximal density ($\rho=1$)
of a compact cluster.

Let us assume that the loose cluster becomes stationary after sufficient 
time has passed. Then the hopping probability of all the ants, except 
the leading one, is assumed to be $H$, while that of the leading one 
is $h$ (see Fig.~\ref{fig-loose}). 
The values of $H$ and $h$ have to be determined self-consistently. 
If $f$ is small enough, then $H$ will be close to $Q$
because the gap between ants is quite small. On the other 
hand, if the density of ants 
is low enough, then $h$ will be very close to $q$ because 
the pheromone dropped by the leading ant would evaporate
when the following ant arrives there.

The typical size of the gap between successive 
ants in the cluster can be estimated \cite{ncs} by considering a 
simple time evolution beginning
with a usual compact cluster (with local density $\rho=1$). 
The leading ant will move forward by one site over the  
time interval $1/h$. This hopping occurs repeatedly and in the 
interval of the successive hopping, the number of the following ants
which will move one step is $H/h$. Thus, in the stationary state, 
strings (compact clusters) of length $H/h$, separated from each other 
by one vacant site, will produced repeatedly by the ants 
(see Fig.~\ref{fig-loose}). 
Then the average gap between ants is 
$\frac{\left(H/h-1\right)\cdot 0+ 1\cdot 1}{H/h}=\frac{h}{H}$,
independent of their density $\rho$. Interestingly, the 
density-independent average gap in the LCA is consistent with the 
flat part (i.e., region~1) observed in computer simulations 
(Fig.~\ref{fig-lca}).
In other words, region~1 is dominated by loose clusters. 

Beyond region 1, the effect of pheromone of the last ant becomes
dominant.  Then the hopping probability of leading ants becomes large
and the gap becomes wider, which will increase the flow.  We call this
region as region 2, in which ``looser'' clusters are formed in the
stationary state. It can be characterized by a negative gradient of
the density dependence of the probability to find an ant in front
of a cell occupied by an ant (see Fig.~\ref{fig-lca}).

Considering these facts, we 
obtain the following equations for $h$ and $H$: 
\begin{eqnarray}
 \left(\frac{h-q}{Q-q}\right)^h=(1-f)^{L-l},\,\,\quad
 \left(\frac{H-q}{Q-q}\right)^{H}=(1-f)^{\frac{h}{H}},
\label{simu}
\end{eqnarray}
where $l$ is the length of the cluster given by
$l=\rho L +(\rho L -1)\frac{h}{H}$.
These equations can be applied to the region 1 and 2.

The total flux in this system is then calculated as follows.
The effective density $\rho_{{\rm eff}}$ in the loose cluster is given by
$\rho_{{\rm eff}}=\frac{1}{1+h/H}$.
Therefore, considering the fact that there are no ants in the part of 
the length $L-l$, the total flux $F$ is
\begin{equation}
 F=\frac{l}{L}f(H,\rho_{{\rm eff}}),
\end{equation}
where $f(H,\rho_{{\rm eff}})$ is given by
\begin{equation}
 f(H,\rho_{{\rm eff}})=\frac12\left(1-\sqrt{1-4H\rho_{{\rm eff}}
(1-\rho_{{\rm eff}})}\right).
\end{equation}

Above the density $\rho=1/2$, ants are assumed to be uniformly distributed
such that a mean-field approximation works well \cite{ncs}.
We call this region as region~3. Thus we have three typical
regions in this model.
In region 3, the relation $H=h$ holds because all the gaps have
the same length, i.e.\ the state is homogeneous. Thus $h$ is determined by
\begin{eqnarray}
 \left(\frac{h-q}{Q-q}\right)^h&=&(1-f)^{\frac1\rho-1},
\label{old}
\end{eqnarray}
which is the same as our previous paper, and
flux is given by $f(h,\rho)$.
It is noted that if we put $\rho=1/2$ and $H=h$, then (\ref{simu})
coincides with (\ref{old}).

In region 1 we can simplify the analysis by assuming $h=q$ in (\ref{simu}). 
Then the flux-density relation becomes linear.
Numerical results for regions 1 and 2 can be obtained by solving
(\ref{simu}) using the Newton method for densities $\rho\le 1/2$. 
Above this value of density, equation (\ref{old}) can be used. 
Together these approximations can reproduce the results of the
simulations rather well \cite{ncs}.

\begin{figure}[tb]
\begin{center}
\ \phantom{a}\\[0.2cm]
\includegraphics[width=0.45\textwidth]{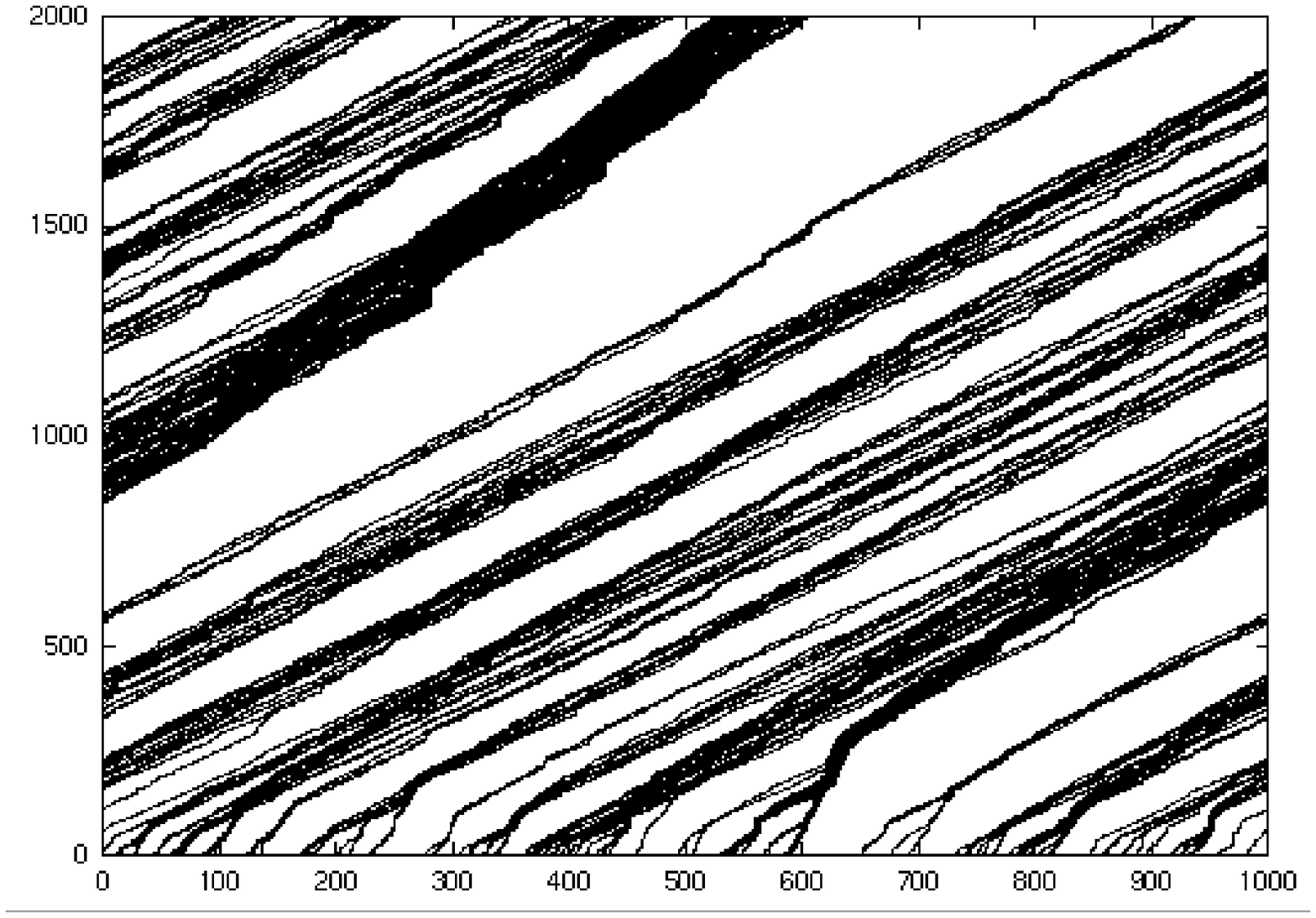}
\qquad
\includegraphics[width=0.45\textwidth]{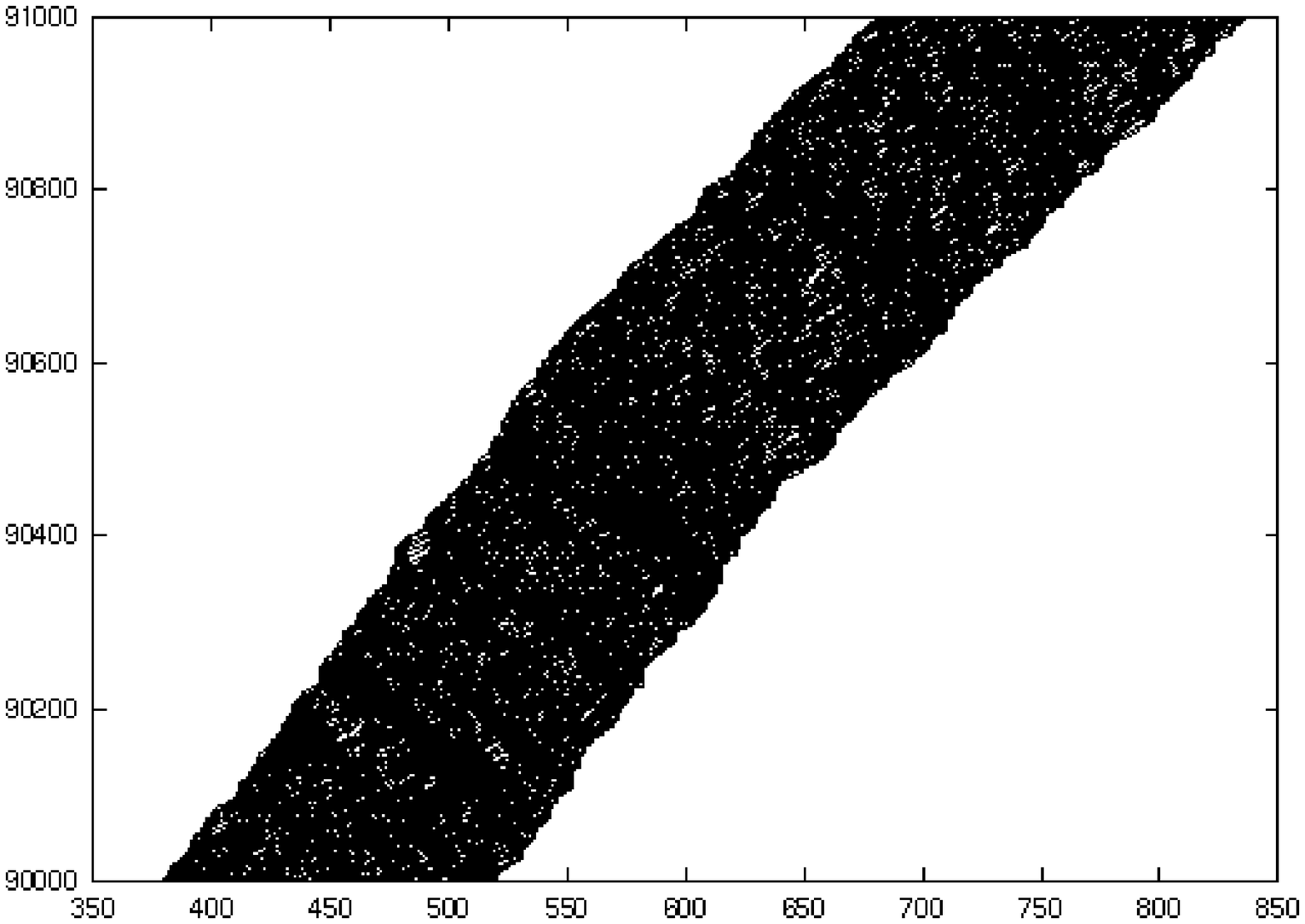}
\end{center}
\caption{Space-time plots showing the coarsening process of the loose 
clusters. The left part shows the state of a system of $L=1000$
cells and $N=100$ ants at early times. 
The right part corresponds to the stationary
state. The evaporation probability of the pheromones is $f=0.0005$.
}
\label{coarse}
\end{figure}
Another interesting phenomenon observed in the simulations is coarsening.
At intermediate time usually several loose clusters are formed
(Fig.~\ref{coarse}). However, the velocity of a cluster depends
on the distance to the next cluster ahead. Obviously, the probability 
that the pheromone created by the last ant of the previous cluster
survives decreases with increasing distance. Therefore clusters with
a small headway move faster than those with a large headway.
This induces a coarsening process such that after long times only
one loose cluster survives (Fig.~\ref{coarse}). A similar
behaviour has been observed in the bus-route model 
\cite{busroute1,busroute2}.


\section{Bidirectional motion}
\label{sec_bidir}

\begin{figure}[tb]
\begin{center}
\includegraphics[width=0.4\textwidth]{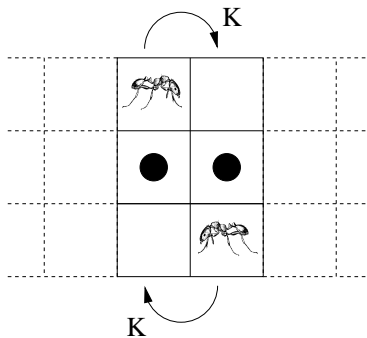}
\end{center}
\caption{A typical head-on encounter of two oppositely moving ants 
in the model of {\it bi-directional} ant-traffic.
This is a totally new process which does not have any analog in the 
model of uni-directional ant-traffic.   
}
\label{fig-bi}
\end{figure}

We develope a model of bi-directional ant-traffic \cite{jscn}
by extending the 
model of uni-directional ant-traffic described in Sec.~\ref{sec_model}. 
In the models of bi-directional ant-traffic the trail consists of 
{\it two} lanes of cells (see Fig.~\ref{fig-bi}). These two lanes need 
not be physically separate rigid lanes in real space; these are, 
however, convenient for describing the movements of ants in two 
opposite directions. In the initial configuration, a randomly selected 
subset of the ants move in the clockwise direction in one lane while 
the others move counterclockwise in the other lane. However, ants are 
allowed neither to take U-turn\footnote{U-turns of so-called
followers on pre-existing trails are very rare \cite{beckers}.} 
nor to change lane. Thus, 
the ratio of the populations of clockwise-moving and anti-clockwise 
moving ants remains unchanged as the system evolves with time. All 
the numerical data presented here are for the {\it symmetric} 
case where equal number of ants move in the two directions. Therefore, 
the {\it average} flux of outbound and nestbound ants are identical. 
In all the graphs we plot only the flux of the nestbound ants.

The rules governing the dropping and evaporation of pheromone in the 
model of bi-directional ant-traffic are identical to those in the 
model of uni-directional traffic. The {\it common} pheromone trail is 
created and reinforced by both the outbound and nestbound ants. The 
probabilities of forward movement 
of the ants in the model of bi-directional ant-traffic are also natural 
extensions of the similar situations in the uni-directional traffic. 
When an ant (in either of the two lanes) {\it does not} face any other 
ant approaching it from the opposite direction the likelihood of its 
forward movement onto the ant-free cell immediately in front of it is 
$Q$ or $q$, respectively, depending on whether or not it finds pheromone 
ahead. Finally, if an ant finds another oncoming ant just in front 
of it, as shown in Fig.~\ref{fig-bi}, it moves forward onto the next 
cell with probability $K$.

Since ants do not segregate in perfectly well defined lanes, head-on 
encounters of oppositely moving individuals occur quite often although 
the frequency of such encounters and the lane discipline varies from 
one species of ants to another. In reality, two ants approaching each 
other feel the hindrance, turn by a small angle to avoid head-on 
collision \cite{couzin} and, eventually, pass each other. 
At first sight, it may appear that the ants in our model follow perfect 
lane discipline and, hence, unrealistic. However, that is not true. 
The violation of lane discipline and head-on encounters 
of oppositely moving ants is captured, effectively, in an indirect 
manner by assuming $K < Q$. But, a left-moving (right-moving) ant 
{\it cannot} overtake another left-moving (right-moving) ant immediately 
in front of it in the same lane. It is worth mentioning that even 
in the limit $K = Q$ the traffic dynamics on the two lanes would 
remain coupled because the pheromone dropped by the outbound ants also 
influence the nestbound ants and vice versa. 

\begin{figure}[tb]
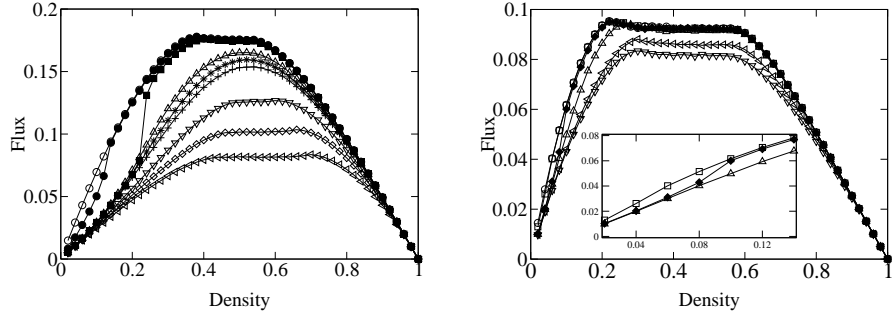

\begin{center}
\ \phantom{a}\\[0.2cm]
\includegraphics[width=0.45\textwidth]{schadfig7a.eps}
\qquad
\includegraphics[width=0.45\textwidth]{schadfig7b.eps}
\end{center}
\caption{Fundamental diagrams of the model for bi-directional traffic 
for the cases $q<K<Q$ (left) and $K<q<Q$ (right) for several different 
values of the pheromone evaporation probability $f$. 
The parameters in the left graph are $Q=0.75, q = 0.25$ and $K=0.5$.
The symbols  $\circ$, $\bullet$, $\blacksquare$, 
$\bigtriangleup$, $\ast$, $+$, $\bigtriangledown$, $\Diamond$  
and $\triangleleft$ correspond, respectively, to 
$f = 0, 0.0005, 0.005,0.05, 0.075,0.10,0.25,0.5$ and $1$.  
The parameters in the right graph are $Q=0.75, q = 0.50$ and $K=0.25$. 
The symbols $\circ$, $\square$, $\blacklozenge$, $\bigtriangleup$, 
$\triangleleft$ and $\bigtriangledown$ correspond, respectively, to 
$f = 0, 0.0005, 0.005, 0.05, 0.5$ and $1$. 
The inset in the right graph is a magnified re-plot of the same data, 
over a narrow range of density, to emphasize 
the fact that the unusual trend of variation of flux with density in 
this case is similar to that observed in the case $q<K<Q$ (left).
The lines are merely guides to the eye. In all cases curves plotted
with filled symbols exhibit non-monotonic behaviour in the speed-density
relation.
}
\label{fig-bifd1}
\end{figure}

Fig.~\ref{fig-bifd1} shows fundamental diagrams for the two relevant
cases $q<K<Q$ and $K<q<Q$ and different values of the evaporation
probability $f$. In both cases the unusual behaviour related to
a non-monotonic variation of the average speed with density
as in the uni-directional model can be observed \cite{jscn}.

An additional feature of the fundamental diagram in the 
bi-directional ant-traffic model is the occurrence of a plateau region. 
This plateau formation is more pronounced in the case $K<q<Q$ than 
for $q<K<Q$ since they appear for all values of $f$.
Similar plateaus have been observed earlier \cite{Janowski,Tripathy} 
in models related to vehicular traffic where randomly placed bottlenecks 
slow down the traffic in certain locations along the route. Note that 
for $q<K<Q$ (Fig.~\ref{fig-bifd1}(left)) the plateaus appear only in the 
two limits $f \rightarrow 0$ and $f \rightarrow 1$, but not for an 
intermediate range of values of $f$. In the limit $f \rightarrow 0$, most 
often the likelihood of the forward movement of the ants is $Q$ 
whereas they are forced to move with a smaller probability $K$ at those 
locations where they face another ant immediately in front approaching 
from the opposite direction (like the situations depicted in 
Fig.~\ref{fig-bi}). 
Thus, such encounters of oppositely moving ants have the same effect on 
ant-traffic as bottlenecks on vehicular traffic.

But why do the plateaus re-appear for $q<K<Q$ also in the limit 
$f \rightarrow 1$? At sufficiently high densities, oppositely moving ants 
facing each other move with probability $K$ rather than $q$. In this case, 
locations where the ants have to move with the lower probability $q$ will 
be, effectively bottlenecks and hence the re-appearance of the plateau. 
As $f$ approaches unity there will be larger number of such locations and, 
hence, the wider will be the plateau. This is consistent with our 
observation in Fig.~\ref{fig-bifd1}(left).


\section{Conclusions}
\label{sec_concl}
We have introduced a stochastic cellular automaton model of an ant trail
\cite{cgns} characterized by two coupled dynamical variables,
representing the ants and the pheromone. With periodic boundary
conditions, the total number of ants is conserved whereas the
total number of pheromones is not conserved. The coupling leads to
surprising results, especially an anomalous fundamental diagram.
This anomalous shape of the fundamental diagram is a consequence of 
the non-monotonic variation of the average speed of the ants with 
their density in an intermediate range of the rate of pheromone 
evaporation. 

As the reason for this unusual behaviour we have identified the
special spatio-temporal organization of the ants and pheromone in the 
stationary state. 
Three different regimes of density can be distinguished by 
studying appropriate correlation functions. At low densities (region 1)
the behaviour is dominated by the existence of loose clusters
which are formed through the interplay between the dynamics of
ants and pheromone. These loose clusters are regions with a density
that is larger than the average density $\rho$, but not maximal.
In region 2, occuring at intermediate densities,
the enhancement of the hopping probability due to pheromone is
dominant. 
Finally, in region 3, at large densities the mutual hindrance against 
the movements of the ants dominates the flow behaviour leading to a 
homogeneous state similar to that of the NaSch model.

Furthermore we have introduced a model of {\it bi-directional} 
ant-traffic. The two main theoretical predictions of this model 
are as follows:\\
(i) The average speed of the ants varies {\it non-monotonically} 
with their density over a wide range of rates of pheromone evaporation. 
This unusual variation gives rise to the uncommon shape of the 
flux-versus-density diagrams. \\
(ii) Over some regions of parameter space, the flux exhibits 
plateaus when plotted against density.

In principle, it should be possible to test these theoretical 
predictions experimentally. Interestingly, various aspects of locomotion 
of {\em individual} ants have been studied in quite great detail 
\cite{lighton,zoll,Weier}. However, traffic is a {\em collective} 
phenomenon involving a large number of interacting ants. Surprisingly, 
to our knowledge, the results published by Burd 
et al.\ \cite{burd} on the leaf-cutting ant {\em Atta Cephalotes} are 
the only set of experimental data available on the fundamental diagram. 
Unfortunately, the fluctuations in the data 
are too high to make any direct comparison with our theoretical 
predictions. Nevertheless it should be possible to observe the
predicted non-monotonicity in future experiments. A simple estimate
shows that the typical magnitudes of $f$, for which the non-monotonic 
variation of the average speed with density is predicted, correspond 
to pheromone lifetimes in the range from few minutes to tens of minutes. 
This is of the same order of magnitude as the measured lifetimes
of real ant pheromones!

\vspace{0.5cm}
\noindent{\bf Acknowledgments}\\
The work of DC is supported, in part, by DFG through a joint
Indo-German research grant.


%

\end{document}